\documentclass{article}
\usepackage{icrctc07}

\title{Selection and reconstruction of very inclined air showers with the Surface Detector of the Pierre Auger Observatory}
\shorttitle{Selection and reconstruction of very inclined air showers}

\authors{D.~Newton$^{1}$ for the Pierre Auger Collaboration$^{2}$}
\shortauthors{D.~Newton}
\afiliations{$^{1}$Departamento de F\'{\i}sica de Part\'{\i}culas, Universidad de Santiago de Compostela, Spain \\  $^{2}$Av. San Martin Norte 304, (5613) Malarg\"{u}e, Prov. De Mendoza, Argentina}
\email{dnewton@fpmacth1.usc.es}

\abstract{The water-Cherenkov tanks of the Pierre Auger Observatory can detect particles
at all zenith angles and are therefore well-suited for the study of inclined
and horizontal air showers ($60^\circ < \theta < 90^\circ$). Such showers are characterised
by a dominance of the muonic component at ground, and by a very elongated and
asymmetrical footprint which can even exhibit a lobular structure due to the
bending action of the geomagnetic field. Dedicated algorithms for the
selection and reconstruction of such events, as well as the corresponding
acceptance calculation, have been set up on basis of muon maps obtained from
shower simulations.}

\begin{document}
\maketitle
%Begin the section.

\section{Introduction}
The Pierre Auger Observatory \cite{eapaper} is sensitive to inclined
 extensive air showers, at  energies above $\sim 5.10^{18}\,\rm{eV}$, with a
 high efficiency and unprecedented statistical accuracy. The characteristics
 of inclined showers, observed with a surface array of detectors,  are quite different to their vertical counterparts and
 require different reconstruction techniques and these characteristics
 offer new  opportunities to  study the properties of the
 primary cosmic ray.

 A  cosmic ray, typically,  initiates an air shower within the first few
 hundred  grams of atmosphere, achieving  shower maximum at $\sim
 800\,\rm{g\,cm^{-2}}$. In the case of vertical showers this  results in a
 large  electro-magnetic component at the ground. 
 Beyond $60^\circ$ ,  the atmospheric  slant depth increases  from
 $1,740\,\rm{g\,cm^{-2}}$,  to $\sim 31,000 \,\rm{g\,cm^{-2}}$ at $90^\circ$ at
 the altitude of the Auger array, and  the  electro-magnetic component of the
 shower is rapidly absorbed -  although below $\sim  65^\circ$ a  significant
 fraction survives at the  ground. Once the primary electro-magnetic
 component has been absorbed, the muons which arrive at the ground are
 accompanied  only by an electro-magnetic halo due, mainly,  to muon decay which
contributes $\sim 15\%$ of the total signal in an Auger surface detector. This
 absorption of the electro-magnetic component significantly affects the Lateral Distribution Function (LDF) of particles, which is used to
 measure the size of vertical air showers, and this makes the vertical  reconstruction algorithm unsuitable for analysing inclined showers. Instead
maps of the muon ground density, based on simulations,  are used to fit the
core location of the shower and the total number of muons.  For highly inclined
showers  the path length of the muons is sufficiently
large that the geomagnetic field significantly affects the muon distribution
on the ground, separating the positive and negative muons and forming a
lobed structure which is replicated in the muon maps. With the aid of the
 maps, the 'size parameter',  $N_{19}$, is measured for each shower. $N_{19}$
 gives the total number of muons,  relative to a shower initiated by a proton
 primary with an energy of $10^{19}\,\rm{eV}$.  The hybrid capability of the
 Auger Observatory (using events observed simultaneously with the
surface array and fluorescence detectors), allows an independent cross-check
 of the geometrical reconstruction algorithm, and also  allows the
 relationship between $N_{19}$ and the energy of the primary particle to be measured \cite{pedro}. 

The inclined
 shower reconstruction algorithm has been developed to  select genuine events
 from the background of atmospheric muons, and to provide a robust measurement
 of the arrival direction and size of the air shower.

\begin{figure}[t]
\centerline{
\includegraphics[width=7cm,height=5.5cm]{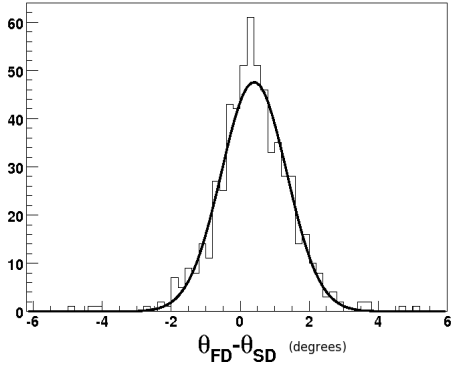}}
\caption{A comparison of the reconstructed zenith angle, $\theta $,
  found using the horizontal reconstruction and the hybrid reconstruction
  \cite{hybrid} for 596 hybrid events. The residuals have a mean of
  $0.45^{\circ}$ and a spread of $0.95^{\circ}$.}\label{fig:angrec}
\end{figure}

\section{Event Selection}

 The trigger hierarchy, for the horizontal reconstruction, follows an
 identical  format to that chosen for the vertical reconstruction
 \cite{trigger}. The Central Trigger (T3) records all candidate events, the Physics Trigger (T4) selects stations which are
 compatible with a shower front moving at the speed of light, and a Quality Trigger (T5)
 is applied, to ensure the validity  of the reconstruction. 

 The first step is to  select physical shower events (T4) from
 the N stations with a signal that were identified by the Central Trigger
 (T3). The timing of each triggered station is checked for compatibility with
 a shower front and the projection of the footprint on the ground plane is
 required to be compact. These tests are applied to an initial
configuration of N selected stations, then successive
trials with N-1, N-2, ... stations are performed until a satisfactory
configuration with four or more stations is found. The conditions to
accept a configuration depend on its zenith angle (the shower front is better
 defined at large zenith angles) and the  multiplicity (the variance of the
 start times increases with the distance to the core).

\begin{figure}[t]
\centerline{
\includegraphics[width=7cm,height=5.5cm]{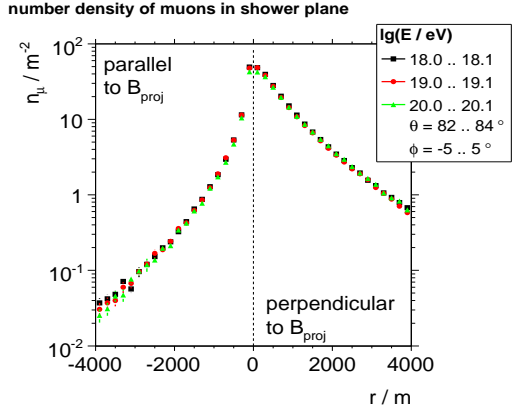}}
\caption{Averaged number density of muons measured in the shower plane. For each primary energy, ten
  CORSIKA proton showers were used. All individual showers were scaled to the equivalent density of $10^{19}\,\rm{eV}$. The
  right (left) hand curve shows the distribution perpendicular (parallel) to
  the projected magnetic field.}\label{fig:energyScaling}
\end{figure}

\section{Angular Reconstruction}

Initially the station start times are corrected for differences in the station
altitude and to compensate for the curvature of the Earth. This gives a  significant
improvement for very inclined, highly energetic events. Of the stations
selected by the T4 trigger, no more than seven (those with the highest signals),
are used in a plane front fit to the corrected start times. The result of this fit
is used to select the appropriate muon map,  which are produced in discrete
steps of $2^{\circ}$ in zenith angle and $5^{\circ}$ in azimuth. With this map the core
location and shower size, $N_{19}$, are provisionally determined (section \ref{maps}).
Once an assumed core location has been found, a more sophisticated angular
reconstruction is made which includes timing corrections to describe the
curvature of the shower front. 
%These corrections are the result of a model
%\cite{time} which takes, as an input,  the distribution of muon production
%heights from Monte Carlo shower simulations, and by considering both geometric
%and kinematical effects, calculates the mean time delay to any point on the
%ground, along with the statistical (sampling) uncertainty.
 The result
from this angular reconstruction, is compared with the result of the original plane fit, and if necessary a more appropriate muon map is selected, and the
angular reconstruction is re-iterated with the new muon map. This process is
repeated until the results converge (typically one iteration is
sufficient).  A comparison of the reconstructed zenith angles  with the hybrid reconstruction
for  596 events is shown in figure \ref{fig:angrec}. 

\section{Core Location and Size Determination with Muon Maps}\label{maps}

Inclined showers have a broken radial symmetry caused primarily by the muon
component, which is long-lived and is polarised by the Earth's magnetic
field. A generalised lateral distribution function is
used to reconstruct these showers, which includes magnetic field
effects. Such a function can be studied and derived from Monte Carlo
simulations to model the lateral number densities of the muons at the
ground. These parameterisations are called muon maps. 

The shape of the muon maps are dependent on zenith and azimuth angle only,
with no significant dependence on the energy  (figure \ref{fig:energyScaling}) and composition of the primary
particle . This invariance  is due, in part, to their strong dependence on the
shape of the muons energy distribution, at their production point, coupled with
the large distance from this production point to the ground. Once the
muons are produced their trajectory to the ground can be described by
well-understood physical processes. 

To derive the muon maps from the Monte Carlo simulations three independent
algorithms were developed, all using proton showers. The different methods
involved AIRES and CORSIKA simulations (using both QGSJET I and QGSJET II), with and without the
geo-magnetic field. For the studies without the magnetic field, geo-magnetic  corrections
were applied using one model which tracks the muons from their production point
to the ground, and a second  model which applies a correction to the ground
distributions.  The resultant ground densities, for each of the three methods,
were then parameterised to produce the final muon maps. An analysis of the
different muon maps  shows a good agreement between all three, with the 
differences far smaller than the Poisson fluctuations expected for a shower
initiated by a primary of energy $10^{19}\,\rm{eV}$. The $sin^2 \theta$
distribution \cite{pedro} for reconstructed events, suggests there is no significant
zenith-dependent bias between the muon maps and the data.

Once the arrival direction has been determined, the shower size and core
reconstruction proceeds. All the selected stations with a signal are used as well as
adjacent stations without a signal. To allow a comparison of the  muon maps
with the station  signals, the signal measured in each tank must be
converted into an equivalent number of muons. As a first step, a correction is
made to remove the  fraction of the signal due to the electro-magnetic
component. This correction is based on a study of AIRES simulations, where the ratio
of the electro-magnetic signal to the muonic signal has been parameterised as
a function of  zenith angle and core distance.  This ratio tends
towards $\sim15\%$ at large core distances and zenith angles. For any assumed core
position the muon maps can be used to predict the number of muons, $N_\mu$, in each
tank. The probability density functions (PDF) for an observed signal, $S$, are
calculated,  based on Geant4 simulations \cite{sims} and take into account the
shower zenith angle and the mean expected muon energy \cite{mag} as well as
the number of muons crossing the tank. Finally the differences between the
muon maps and the corrected station signals are minimised to find the core location and $N_{19}$, the number of muons in the shower, relative to the appropriate muon map. 

\section{Quality Trigger (T5) and Aperture Calculation}
 Following the strategy used by the Pierre Auger Collaboration to
produce the spectrum for vertical showers \cite{ap}, the
acceptance is calculated geometrically. Two basic T5 configurations are considered:
1) The station closest to the reconstructed core location  must be surrounded by an `active' hexagon  (i.e. six functioning
adjacent stations, though not necessarily with a signal) and 2) the station
closest to the core must be surrounded by two `active' hexagons (18
functioning stations). In addition the reconstructed core location must be
enclosed by a triangle of active stations. Compromises on these criteria  are also being
considered, allowing for one missing station in the hexagons.

The geometric computation is based on counting the stations 
that fulfill the requirement imposed by the T5 quality trigger.  Moreover, the T3 central trigger 
condition must be fulfilled by stations involved in the T5 to ensure a
uniform response from the array. The central trigger assesses up to four hexagons of stations surrounding the central station to built a T3. For inclined
showers, which do not have a compact configuration on the detector plane, the
energy at which  the T3 efficiency reaches $100\%$ will increase if the
T3 condition is required in fewer than four hexagons. If  more active hexagons
are required in the T5 trigger, the acceptance decreases. With two active 
hexagons, it decreases to $\sim 50\%$ ($\sim 80 \%$ allowing one missing station) of that with one hexagon.  A comparison is also underway to compute the acceptance by Monte Carlo,
   throwing simulated muon maps at a realistic detector array, using a  position of
   the core and a time of occurrence that are randomly selected.  This avoids the
   compromise, between maximising the acceptance and reducing the energy at which the efficiency of the array is $100\%$, that appears
   for the geometric computation.

The quality of the
reconstruction is currently being assessed under the various T5
 conditions. This is done using real showers, hitting the centre
 of an ideal array in which specific real configurations with holes and edges
 are forced. 
Preliminary results suggest the dispersion on the
reconstructed size parameter is negligible  with the requirement that the closest
station to the shower core is surrounded by six active stations. Other configurations with requirements for the next
to closest neighbours do not significantly reduce the dispersion of the
reconstructed size parameter.

\section{Outlook}

The signals, measured with the surface array, from inclined showers lead to what
is essentially a measurement of the muon content of the shower ($N_{19}$). Combined with
measurements of the electro-magnetic content and the depth of shower maximum, (with the Auger fluorescence detector), this gives a powerful
 tool to study the cosmic ray composition \cite{comp}. Additionally the detectors are
 sensitive to both deeply-interacting, and Earth-skimming, inclined neutrinos which
 can be discriminated from the nucleonic cosmic ray flux \cite{neu1}. Analysing inclined
 showers increases the Auger aperture significantly: half the available solid
 angle corresponds to zenith angles between $60$ and $90^{\circ}$. The analysis of inclined
 showers will offer insights into the cosmic ray
 composition and their atmospheric interactions, and will also supplement the
 vertical observations by increasing the available number of events in the
 measurement of the  cosmic ray flux and in anisotropy studies.

%This is the reference to .bib file (Without .bib!)
\bibliography{icrc0308}
%This in the bibtex style, is ok.
\bibliographystyle{plain}
\end{document}